\documentclass[3p]{elsarticle}
\usepackage{graphicx}
\usepackage{dcolumn}
\usepackage{bm}
\usepackage{epsfig}
\usepackage{subfigure}
\usepackage{multirow}
\usepackage{amsmath} 
\usepackage{hyperref}
\journal{Physics Letters B}

\begin{document}

\begin{frontmatter}

\title{A new texture of neutrino mass matrix with three constraints}

\author[1]{Radha Raman Gautam}
\ead{gautamrrg@gmail.com}

\author[2]{Sanjeev Kumar}
\ead{skverma@physics.du.ac.in}

\address[1]{Department of Physics, LBS Govt. PG College, Saraswatinagar, Shimla-171206, India}
\address[2]{Department of Physics and Astrophysics, University of Delhi, Delhi-110007, India}

\begin{abstract}
We present a new texture of neutrino mass matrix having three complex relations among its elements and study in detail the phenomenological implications. A characteristic feature of the resulting neutrino mass matrix is that the atmospheric neutrino mixing angle is predicted to lie in a very narrow region near $45^{\circ}$. We illustrate how such a form of the neutrino mass matrix can be realized using the non-Abelian flavor symmetry $A_4$ in the framework of type-I+II seesaw mechanism.
\end{abstract}
\end{frontmatter}

\section{Introduction}
The neutrino experiments aim to reconstruct the neutrino mass matrix which contains information about neutrino masses, mixing angles and CP-violating phases. We now have rather accurate measurements of the three mixing angles ($\theta_{12},\theta_{13},\theta_{23}$) and two mass squared differences ($\Delta m_{21}^2,\Delta m_{23}^2$). However, the octant of $\theta_{23}$, sign of $\Delta m_{23}^2$, and CP-violating phases are yet to be determined. The direct and cosmological searches of neutrino mass have considerably reduced the upper bound on the absolute neutrino masses scale. In order to determine the neutrino mass matrix, we need the full knowledge of all the nine physical observables: three neutrino masses, three mixing angles and three CP-violating phases.

Even with the incomplete knowledge of these neutrino parameters, we can still obtain valuable information about the possible structure of the neutrino mass matrix. For example, the not so small value of $\theta_{13}$ disfavours Tribimaximal (TBM) \cite{tbm}. However, the variations of TBM like TM1 and TM2 mixings are still allowed \cite{He,WR}. There are many other structures of the neutrino mass matrix like texture zeros \cite{fgm,tz},  vanishing cofactors \cite{cofactor}, equalities \cite{twoequality}, hybrid textures \cite{hybrid} that are consistent with the current neutrino oscillation data.

One can construct models for specific structures of neutrino mass matrix by incorporating flavor symmetries in the standard model (SM) of particle Physics. The first attempt to accommodate TBM mixing scheme in a neutrino mass model was  based upon the non-Abelian discrete symmetry group $A_4$ \cite{a4}. Many other non-Abelian discrete symmetry groups were later used to construct neutrino mass models with TBM mixing \cite{discrete}. The observation of non-zero $\theta_{13}$ lead to modifications of such models \cite{partial}. 

Other successful attempts to constrain the structure of neutrino mass matrix are texture zeros and vanishing cofactors. These patters are generally realized using Abelian discrete flavor symmetries. While TBM mixing and its variants predicted mixing angles or relations among them, the texture zeros, vanishing cofactors, and equalities predict relations between neutrino masses, mixing angles, and CP violating phases. Another approach combines variants of TBM mixing
with texture zeros/vanishing cofactors \cite{ourtm}.
\section{Neutrino Mass Matrix}
In the present work we propose the following texture of the neutrino mass matrix in the flavor basis:
\begin{equation}
 M_\nu = \left(
\begin{array}{ccc}
x & y & -y \\ y & 2 y  & z \\ -y & z & -2 y
\end{array}
\right).
\end{equation}
The number of free parameters in the above neutrino mass matrix are reduced from 12 to 6 by imposing three independent relations among the neutrino mass matrix elements:
\begin{equation}
\label{rel1}
M_{12} = - M_{13}
\end{equation}
 
\begin{equation}
\label{rel2}
M_{22}=-M_{33}
\end{equation}

\begin{equation}
\label{rel3}
M_{22}=2M_{12}
\end{equation}

The fourth relation $M_{33}=2M_{13}$, derivable from the first three, is not independent.

First two relations \ref{rel1} and \ref{rel2} are expressions of $\mu-\tau$ anti-symmetry \cite{antimutau}. But, a mass matrix with $\mu-\tau$ anti-symmetry should also have vanishing $(1,1)$ and $(2,3)$ entries. Since, $\mu-\tau$ anti-symmetry is not allowed experimentally, we need non-zero $M_{11}$ and $M_{23}$ to explain the neutrino masses \cite{antimutau}. The third relation enhances the predictive power of this texture by imposing an additional relation on the mass matrix elements apart from the two relations of the $\mu-\tau$ anti-symmetry.

To analyse the phenomenological implications of present neutrino mass model, we reconstruct the neutrino mass matrix in the flavor basis (i.e. in the basis where the charged lepton mass matrix is diagonal) assuming neutrinos to be Majorana particles. In this basis, a complex symmetric neutrino mass matrix can be diagonalized by a unitary matrix $V'$ as
\begin{equation}
M_{\nu}= V' M_{\nu}^{\textrm{diag}}V'^{T}
\label{eq5}
\end{equation}
where $M_{\nu}^{\textrm{diag}}$ = diag$(m_1,m_2,m_3)$. \\
The unitary matrix $V'$ can be parametrized as
\begin{equation}
V' = P_lV \ \ \ \textrm{with}\ \ \ \ V = UP_\nu
\label{eq6}
\end{equation}
where  \cite{foglipdg}
\begin{equation}
U= \left(
\begin{array}{ccc}
c_{12}c_{13} & s_{12}c_{13} & s_{13}e^{-i\delta} \\
-s_{12}c_{23}-c_{12}s_{23}s_{13}e^{i\delta} &
c_{12}c_{23}-s_{12}s_{23}s_{13}e^{i\delta} & s_{23}c_{13} \\
s_{12}s_{23}-c_{12}c_{23}s_{13}e^{i\delta} &
-c_{12}s_{23}-s_{12}c_{23}s_{13}e^{i\delta} & c_{23}c_{13}
\end{array}
\right)
\end{equation}
with $s_{ij}=\sin\theta_{ij}$ and $c_{ij}=\cos\theta_{ij}$ and
\begin{small}
\begin{center}
$P_\nu = \left(
\begin{array}{ccc}
1 & 0 & 0 \\ 0 & e^{i\alpha} & 0 \\ 0 & 0 & e^{i(\beta+\delta)}
\end{array}
\right)$ , \ \ \ 
$P_l = \left(
\begin{array}{ccc}
e^{i\varphi_e} & 0 & 0 \\ 0 & e^{i\varphi_\mu} & 0 \\ 0 & 0 & e^{i\varphi_\tau}
\end{array}
\right).$
\end{center}
\end{small}
Here, $P_\nu$ is the diagonal phase matrix with two Majorana-type CP- violating phases $\alpha$, $\beta$ and one Dirac-type CP-violating phase $\delta$. The phase matrix $P_l$ contains unphysical phases that depend on the phase convention. The matrix $V$ is called the neutrino mixing matrix or the Pontecorvo-Maki-Nakagawa-Sakata (PMNS) matrix  \cite{pmns}. Using Eqs. (\ref{eq5}) and Eq. (\ref{eq6}), the neutrino mass matrix can be written as
\begin{equation}
M_{\nu}=P_l U P_\nu M_{\nu}^{\textrm{diag}}P_\nu^{T}U^{T}P_l^T.
\end{equation}

The simultaneous existence of relations (\ref{rel1}) and (\ref{rel2}) between the elements of the neutrino mass matrix implies
\begin{align}
& e^{i(\varphi_e +\varphi_\mu)}M_{\nu (e \mu)} + e^{i(\varphi_e +\varphi_\tau)}M_{\nu (e \tau)} = 0\ , \\ & e^{i(\varphi_{\mu}+\varphi_{\mu})} M_{\nu (\mu \mu)} + e^{i(\varphi_{\tau} +\varphi_{\tau})} M_{\nu (\tau \tau)} = 0 \ \ 
\end{align}
or
\begin{align}
& Q M_{\nu (e \mu)} + M_{\nu (e \tau)} = 0\ , \\ & Q' M_{\nu (\mu \mu)} + M_{\nu (\tau \tau)} = 0 
\end{align}
where
\begin{align}
Q = & e^{i(\varphi_e +\varphi_{\mu} - (\varphi_e +\varphi_{\tau}))}\ , \\ Q' = & e^{i(\varphi_{\mu}+\varphi_{\mu}-(\varphi_{\tau} +\varphi_{\tau}))}\ .
\end{align}
In this notation, the relations (\ref{rel1}) and (\ref{rel2}) between the neutrino mass matrix elements yield two complex equations viz.
\begin{align}
\sum_{i=1}^{3} & (Q V_{e i}V_{\mu i} + V_{e i}V_{\tau i})m_i = 0 \ , \\
\sum_{i=1}^{3} & (Q' V_{\mu i}V_{\mu i} + V_{\tau i}V_{\tau i})m_i = 0  \ .
\end{align}
A more elegant version of these equations can be rewritten as
\begin{eqnarray}
m_1 A_1 + m_2 A_2 e^{2i\alpha} + m_3 A_3 e^{2i(\beta +\delta)}=0 \ ,
\label{eq17} \\
m_1 B_1 + m_2 B_2 e^{2i\alpha} + m_3 B_3 e^{2i(\beta +\delta)}=0 
\label{eq18}
\end{eqnarray}
where
\begin{equation}
A_i =(Q U_{e i }U_{\mu i} + U_{e i}U_{\tau i}) \ , \ \ 
B_i =(Q' U_{\mu i}U_{\mu i} + U_{\tau i}U_{\tau i}) \ 
\end{equation}
with $(i = 1, 2, 3)$. 

The two complex Eqs.(\ref{eq17}) and (\ref{eq18}) involve nine physical parameters: $m_{1}$, $m_{2}$, $m_{3}$, $\theta _{12}$, $\theta _{23}$, $\theta _{13}$, and three CP-violating phases ($\alpha $, $\beta $, and $\delta $). In addition, there are three unphysical phases $(\varphi_e, \varphi_\mu, \varphi_\tau)$ which enter in the mass ratios as two phase differences. The masses $m_{2}$ and $m_{3}$ can be calculated from the mass-squared differences $\Delta m_{21}^{2}$ and $|\Delta m_{32}^{2}|$ using the relations
\begin{equation}
m_{2}=\sqrt{m_{1}^{2}+\Delta m_{21}^{2}} \ , \ \  m_{3}=\sqrt{m_{2}^{2} \pm |\Delta m_{32}^{2}|} \ 
\end{equation}
where $m_2 > m_3$ for an inverted neutrino mass ordering (IO) and $m_2 < m_3$ for the normal neutrino mass ordering (NO). Using the experimental inputs of the two mass-squared differences and the three mixing angles we can constrain the other parameters. 

Simultaneously solving Eqs.(\ref{eq17}) and (\ref{eq18}) for two mass ratios, we obtain
\begin{small}
\begin{equation}
\frac{m_1}{m_2}e^{-2i\alpha }=\frac{A_2 B_3 - A_3 B_2}{A_3 B_1 - A_1 B_3}
\label{eq21}
\end{equation}
\end{small}
and
\begin{small}
\begin{equation}
\frac{m_1}{m_3}e^{-2i\beta }=\frac{A_3 B_2 - A_2 B_3 }{A_2 B_1- A_1 B_2}e^{2i\delta } \ .
\label{eq22}
\end{equation}
\end{small}
The magnitudes of the two mass ratios in Eqs.(\ref{eq21}) and (\ref{eq22}) are given by
\begin{equation}
r_{13}=\left|\frac{m_1}{m_3}e^{-2i\beta }\right| ,
\label{eq23}
\end{equation}
\begin{equation}
r_{12}=\left|\frac{m_1}{m_2}e^{-2i\alpha }\right|
\label{eq24}
\end{equation}
while the CP-violating Majorana phases $\alpha$ and $\beta$ are given by
 \begin{small}
\begin{align}
\alpha & =-\frac{1}{2}\textrm{Arg}\left(\frac{A_2 B_3 - A_3 B_2}{A_3 B_1 - A_1 B_3}\right), \\
\beta & =-\frac{1}{2}\textrm{Arg}\left(\frac{A_3 B_2 - A_2 B_3 }{A_2 B_1- A_1 B_2}e^{2i\delta }\right).
\end{align}
\end{small}
As $\Delta m_{21}^{2}$ and $|\Delta m_{32}^{2}|$ are known experimentally, the values of a mass ratio ($r_{13}$ or $r_{12}$) from Eq.(\ref{eq23}) or Eq.(\ref{eq24}) can be used to calculate $m_1$. For example, by inverting Eq.(\ref{eq23}), we obtain
\begin{small}
\begin{equation}
m_{1}=r_{13} \sqrt{\frac{\Delta m_{21}^{2}+
|\Delta m_{32}^{2}|}{ 1-r_{13}^{2}}} .
\end{equation}
\end{small}
Also, the mass ratios in Eqs. (\ref{eq23}) and (\ref{eq24}) can be used to obtain the expression for the parameter $R_\nu$, which we define as the ratio of mass squared differences ($\Delta m_{ij}^2 = m_i^2 - m_j^2$):
\begin{equation}\label{eq:rnu}
R_\nu \equiv \frac{\Delta m_{21}^2}{|\Delta m_{31}^2|} = \frac{(\frac{m_2}{m_1})^2-1}{|(\frac{m_3}{m_1})^2-1|}
\end{equation}
where $m_1 > m_3$ for an IO and $m_1 < m_3$ for the NO.
For Eqs. (\ref{rel1}) and (\ref{rel2}) to be consistent with the present neutrino oscillation data, the parameter $R_\nu$ should lie within its experimentally allowed range. 

We solve the three constraints on the neutrino mass matrix (Eqs. (\ref{rel1})-(\ref{rel3})) by generating sufficiently large number of random points for our free parameters. The points in the parameter space that satisfy the constraints of Eqs. (\ref{rel1}) and (\ref{rel2}) are then required to satisfy the third relation in Eq. (\ref{rel3}) with the identical accuracy level determined by the experimental error on the neutrino oscillation parameters. The current experimental best fit values and the corresponding 1$\sigma$ and 3$\sigma$ errors used in the numerical analysis are given in Table \ref{tabdat}. We also have an upper bound on the sum of neutrino masses:
\begin{equation}
\Sigma = \sum_{i=1,2,3}^{3} m_{i} .
\end{equation} 
Planck satellite data \cite{planck} combined with WMAP, CMB and BAO experiments limit the sum of neutrino masses $\sum m_{i}\leq 0.12$ eV at 95$\%$ confidence level (CL). In the present work, we assume a more conservative limit of $\sum m_{i}\leq 1$ eV.

\begin{figure*}
\centering 
\includegraphics[scale=0.33]{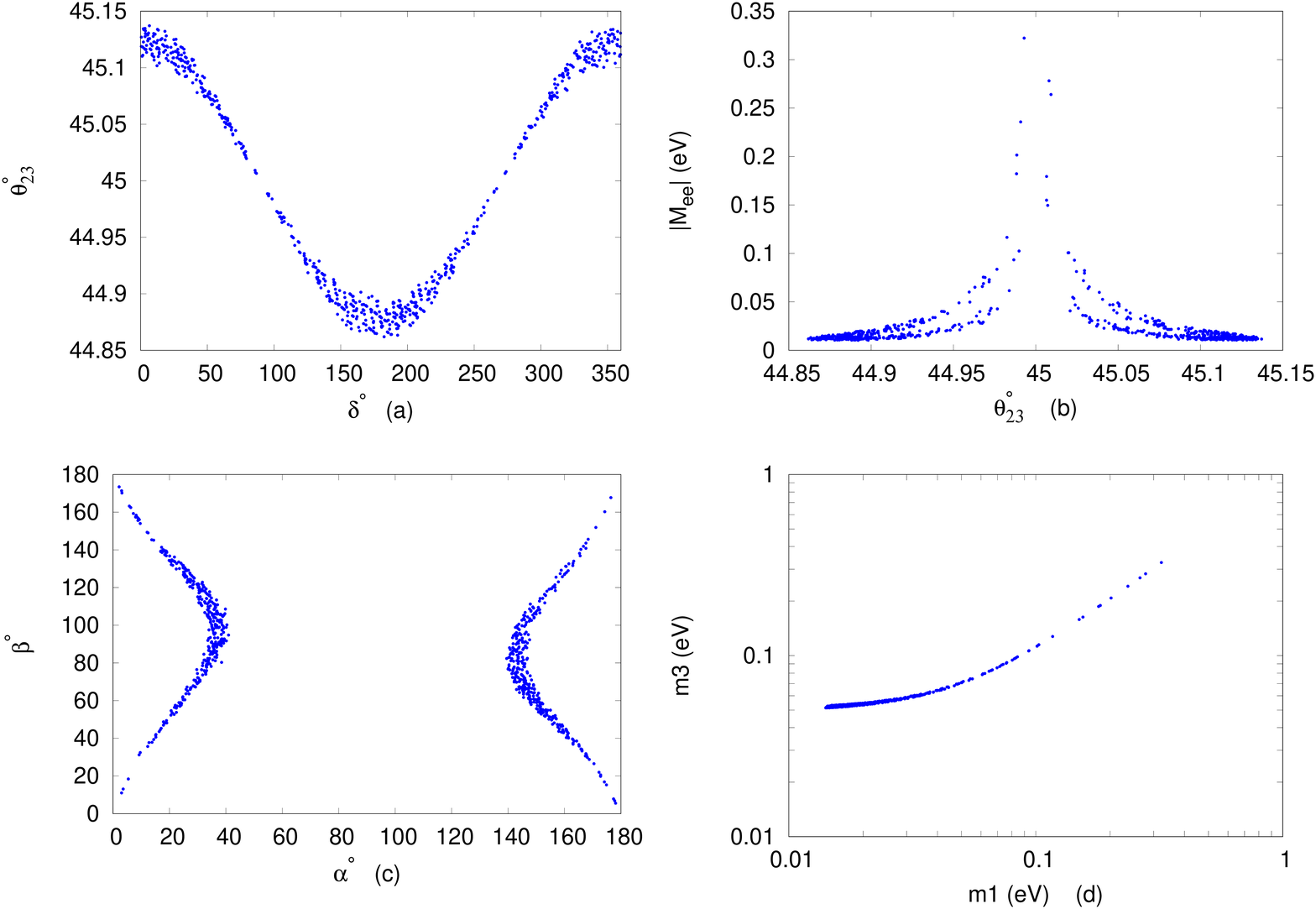}
\caption{Correlation plots for normal neutrino mass ordering}
\label{fig1}
\end{figure*}

\begin{figure*}[t]
\centering 
\includegraphics[scale=0.33]{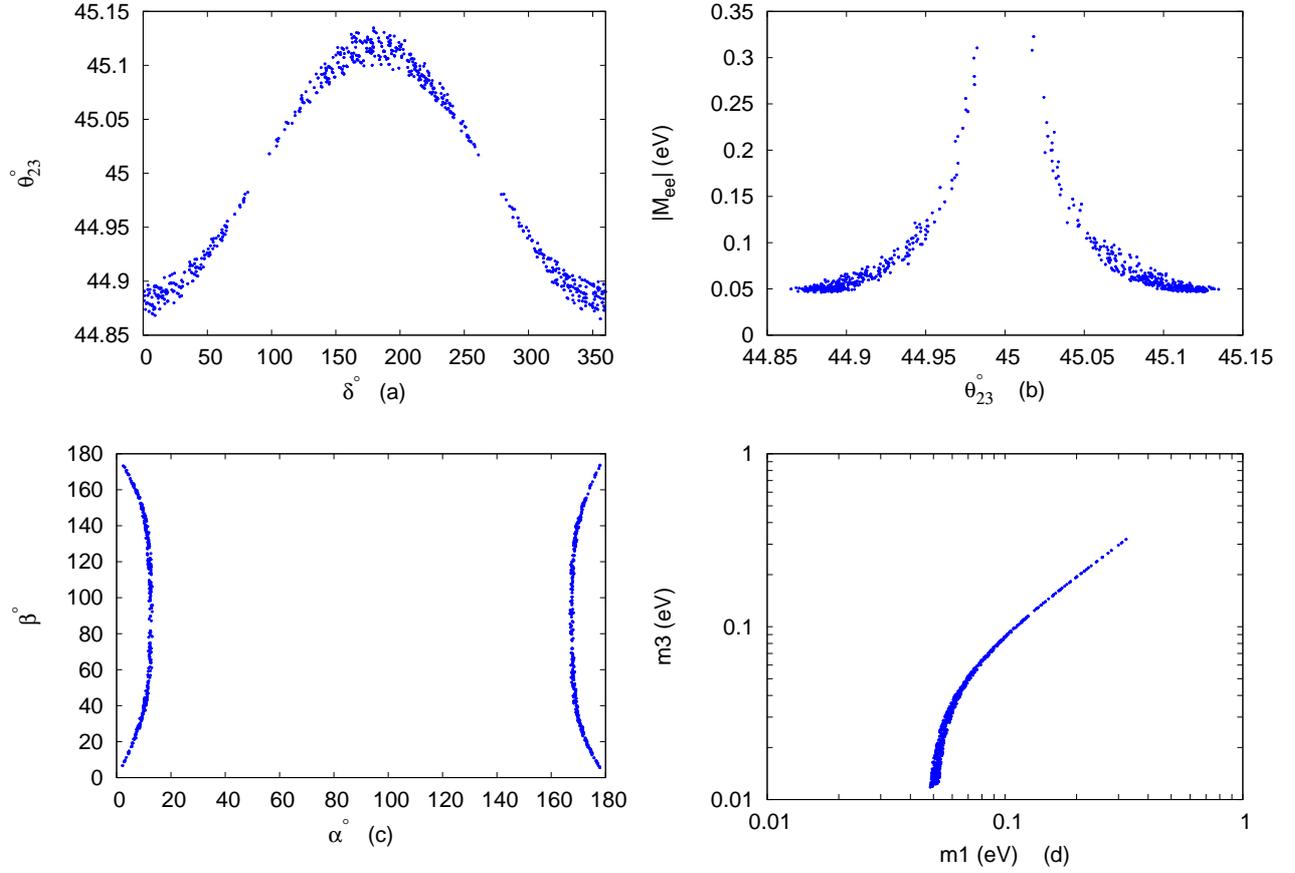}
\caption{Correlation plots for inverted neutrino mass ordering}
\label{fig2}
\end{figure*}

\begin{table*}[t]
\begin{center}
\begin{tabular}{|c|c|c|}
 \hline
Parameter & Normal Ordering & Inverted Ordering \\
  & best fit $\pm 1 \sigma$ ~~ $3 \sigma$ range  & best fit $\pm 1 \sigma$~~~~~~ $3 \sigma$ range \\
 \hline 
$\theta_{12}^{\circ}$ & $33.44^{+0.77}_{-0.74}$ ~ $31.27$ - $35.86$ & $33.45^{+0.78}_{-0.75}$ ~~~ $31.27$ - $35.87$ \\
$\theta_{23}^{\circ}$ & $49.2^{+0.9}_{-1.2}$  ~~$40.1$ - $51.7$ & $49.3^{+0.9}_{-1.1}$ ~~~~~~ $40.3$ - $51.8$ \\
$\theta_{13}^{\circ}$ & $8.57^{+0.12}_{-0.12}$  ~~$8.20$ - $8.93$ & $8.60^{+0.12}_{-0.12}$ ~~~~~~ $8.24$ - $8.96$ \\
$\delta_{CP}^{\circ}$ & $197^{+27}_{-24}$ ~~ $120$ - $369$ &  $282^{+26}_{-30}$ ~~~~~~ $193$ - $352$ \\
$\Delta m^{2}_{21}/10^{-5} eV^2 $ & $7.42^{+0.21}_{-0.20}$ ~~$6.82$ - $8.04$ & $7.42^{+0.21}_{-0.20}$ ~~~~ $6.82$ - $8.04$ \\
$|\Delta m^{2}_{3l}|/10^{-3} eV^2 $ & $2.517^{+0.026}_{-0.028}$ ~~ $2.435$ - $2.598$ & $2.498^{+0.028}_{-0.028}$ ~~~~~~ $2.581$ - $2.414$ \\
 \hline 
 \end{tabular}
\caption{Current neutrino oscillation parameters from global fits \cite{data}. Here $\Delta m_{3l}^2 \equiv \Delta m_{31}^2 > 0$ for normal ordering and $\Delta m_{3l}^2 \equiv \Delta m_{32}^2 < 0$ for inverted ordering}
\label{tabdat}
\end{center}
\end{table*}
The results of this analysis are shown as correlation plots in figures \ref{fig1} and \ref{fig2}. We have depicted the points in the allowed parameter space at the 3$\sigma$ confidence level.  A characteristic feature of this texture is that the atmospheric neutrino mixing angle $\theta_{23}^{\circ}$ is predicted to lie in a very narrow region near $45^{\circ}$ (Fig.\ref{fig1} for NO and Fig.\ref{fig2} for IO). Also, the Dirac-type CP violating phase $\delta$ is correlated with $\theta_{23}^{\circ}$ (Fig. \ref{fig1}(a) for NO and Fig. \ref{fig2}(a) for IO).  As a generic prediction of such textures, the two Majorana phases show sharp correlations with one other (Fig.\ref{fig1}(c) for NO and \ref{fig2}(c) for IO). 

Another notable prediction of this model is a lower bound on the effective mass relevant to neutrinoless double beta decay ($|M_{ee}| > 0.046$  eV) for inverted neutrino mass ordering (Fig.\ref{fig2}(b)) which is testable in the currently running and forthcoming neutrinoless double beta decay experiments \cite{ndb}. 

We also depict the correlation plots of the neutrino masses ($m_1$ and $m_3$) in Fig.\ref{fig1}(d)  (for NO) and Fig. \ref{fig2}(d) (for IO). For the present neutrino mass model, the sum of neutrino masses ($\sum m_i$) remains above for 0.08 eV for NO and 0.1 eV for IO which implies that the neutrino mass matrix realized in present work predicts a quasi-degenerate neutrino mass spectrum. \\
\section{Symmetry Realization}
To obtain the desired form of the form of the neutrino mass matrix from $A_4$ symmetry, we extend the standard model (SM) by adding two additional $SU(2)_L$  doublet Higgs fields to the SM and three $SU(2)_L$ triplet Higgs fields (we do not discuss the Higgs phenomenology in this work). The additional triplet Higgs fields lead to non-zero neutrino masses via the type-II seesaw mechanism \cite{seesaw2}. We also add three right handed neutrino fields that contribute to the effective neutrino mass matrix via the type-I seesaw mechanism \cite{seesaw1}.  The transformation properties of various leptonic and Higgs fields under $A_4$ flavor symmetry and SM gauge group $SU(2)_L$ are given in Table-~\ref{tab:trans}.

\begin{table}
\begin{center}
\begin{tabular}{ccccccccc}
\hline 
Fields & $D_{l_{L}}$ & $l_R $ & $\nu_{l_{R}}$ & $\phi $ & $\Delta $\\ 
\hline 
$SU(2)_L$ & 2 & 1 & 1 & 2 & 3 \\ 
$A_4$ & 3 & ($1,1^{\prime \prime},1^\prime$) & ($1,1^{\prime \prime},1^\prime$) & 3 & 3  \\ 
\hline 
\end{tabular}
\end{center}
\caption{Transformation properties of various fields: $D_{l_{L}}  (D_{e_{L}}, D_{\mu_{L}}, D_{\tau_{L}})^T$, 
$l_R (e_R, \mu_R, \tau_R)^{T}$, $\nu_{l_{R}} (\nu_{e_R}, \nu_{\mu_R}, \nu_{\tau_R})^{T}$,  $\phi (\phi_1, \phi_2, \phi_3)^T$ and 
$\Delta (\Delta_1, \Delta_2, \Delta_3)^T$.}
\label{tab:trans}
\end{table}
The above transformation properties lead to the following $A_4$ invariant Yukawa Lagrangian:
\begin{eqnarray}
-\mathcal{L}_{\text{Yukawa}} & = & y_1 (\overline{D}_{e_L} \phi_1 + \overline{D}_{\mu_L} \phi_2 
+ \overline{D}_{\tau_L} \phi_3)_{\underline{1}} e_{R_{\underline{1}}}   
\nonumber \\
& + &  y_2 (\overline{D}_{e_L} \phi_1 + \omega^2 
\overline{D}_{\mu_L} \phi_2 +  \omega \overline{D}_{\tau_L} \phi_3)_{\underline{1}^\prime} \mu_{R_{\underline{1}^{\prime\prime}}}  
\nonumber \\
& + & y_3 (\overline{D}_{e_L} \phi_1 + \omega \overline{D}_{\mu_L} \phi_2 
+ \omega^2 \overline{D}_{\tau_L} \phi_3)_{\underline{1}^{\prime \prime}} \tau_{R_{\underline{1}^{\prime}}}  \nonumber \\
& + & y_4 (\overline{D}_{e_L} \tilde{\phi}_1 + \overline{D}_{\mu_L} \tilde{\phi}_2 
+ \overline{D}_{\tau_L} \tilde{\phi}_3)_{\underline{1}} \nu_{e_R{\underline{1}}}   
\nonumber \\
& + &  y_5 (\overline{D}_{e_L} \tilde{\phi}_1 + \omega^2 
\overline{D}_{\mu_L} \tilde{\phi}_2 
+  \omega \overline{D}_{\tau_L} \tilde{\phi}_3)_{\underline{1}^\prime} \nu_{\mu_R{\underline{1}^{\prime\prime}}}  \nonumber \\
& + & y_6 (\overline{D}_{e_L} \tilde{\phi}_1 
+ \omega \overline{D}_{\mu_L} \tilde{\phi}_2 
+ \omega^2 \overline{D}_{\tau_L} \tilde{\phi}_3)_{\underline{1}^{\prime \prime}} \nu_{\tau_R{\underline{1}^{\prime}}}  \nonumber \\ 
& - & y_{\Delta} [(D_{\mu_L}^T C^{-1} D_{\tau_L}+D_{\tau_L}^T C^{-1} D_{\mu_L})i \tau_2 \Delta_1 
\nonumber \\
& + & (D_{\tau_L}^T C^{-1} D_{e_L}+D_{e_L}^T C^{-1} D_{\tau_L})i \tau_2 \Delta_2 
\nonumber \\ 
& + & (D_{e_L}^T C^{-1} D_{\mu_L}+D_{\mu_L}^T C^{-1} D_{e_L})i \tau_2 \Delta_3]  
\nonumber \\ 
& - &  M_1 (\nu_{e_R}^T C^{-1} \nu_{e_R} )_{\underline{1}}  
\nonumber \\
& - & M_2 (\nu_{\mu_R}^T C^{-1} \nu_{\tau_R} + \nu_{\tau_R}^T C^{-1} \nu_{\mu_R} )_{\underline{1}} + \ \ \textrm{H.c.}
\end{eqnarray}
where $\tilde{\phi}=i \tau_2 \phi^*$. 

For the $\phi_i$ Higgs fields, the vacuum expectation values (VEVs) along the direction $\langle \phi \rangle_o = v_\phi (1,1,1)^T$ lead to the following charged lepton mass matrix in the symmetry basis:
  \begin{equation}
 m_l = \left(
\begin{array}{ccc}
y_1 v_{\phi} & y_2 v_{\phi} & y_3 v_{\phi} \\ 
y_1 v_{\phi} & y_2 \omega v_{\phi} & y_3 \omega^2 v_{\phi} \\
y_1 v_{\phi} & y_2 \omega^2 v_{\phi} & y_3 \omega v_{\phi}
\end{array}
\right).
 \end{equation}
The Dirac neutrino mass matrix has a structure similar to the charged lepton mass matrix i.e.
\begin{equation}
 m_D = \left(
\begin{array}{ccc}
y_4 v_{\phi} & y_5 v_{\phi} & y_6 v_{\phi} \\ 
y_4 v_{\phi} & y_5 \omega v_{\phi} & y_6 \omega^2 v_{\phi} \\
y_4 v_{\phi} & y_5 \omega^2 v_{\phi} & y_6 \omega v_{\phi}
\end{array}
\right).
 \end{equation} 
 The right-handed neutrino mass matrix has the following form:
 \begin{equation}
 m_R = \left(
\begin{array}{ccc}
M_1 & 0&0\\ 
0 & 0&M_2 \\
0 & M_2 &0
\end{array}
\right).
 \end{equation}
The type-I seesaw contribution to the effective neutrino mass matrix after redefining some parameters takes the following form:
\begin{equation}
 m_{\nu_I} \approx m_D m_R^{-1}m_D^{T} = \left(
\begin{array}{ccc}
a + 2 b & a - b & a - b \\ 
a - b & a + 2 b & a - b \\
a - b & a - b & a + 2 b
\end{array}
\right).
 \end{equation}
 
 For the type-II seesaw contribution to the effective neutrino mass matrix, the $A_4$ and $SU(2)_L$ triplet Higgs fields are assumed to have the VEV alignment $\langle \Delta \rangle_o= v_\Delta (0,-1,1)^T$. Such an alignment of VEVs has been achieved in Refs. \cite{vev1,vev2} by allowing specific terms in the scalar potential which break $A_4$ softly. This leads to the following form of the neutrino mass matrix via the type-II seesaw mechanism:
\begin{equation}
 m_{\nu_{II}} = \left(
\begin{array}{ccc}
0& y_{\Delta} v_{\Delta} &-y_{\Delta} v_{\Delta} \\ y_{\Delta} v_{\Delta} & 0 & 0 \\ -y_{\Delta} v_{\Delta} & 0 & 0
\end{array}
\right),
\end{equation}
 
which can be redefined as:
\begin{equation}
 m_{\nu_{II}} = \left(
\begin{array}{ccc}
0 & c & -c \\ c & 0  & 0 \\ -c & 0 & 0
\end{array}
\right).
\end{equation}

The complete effective neutrino mass matrix in the symmetry basis has the following form:
\begin{equation}
 m_\nu \equiv m_{\nu_I} + m_{\nu_{II}} = \left(
\begin{array}{ccc}
a + 2 b & a - b + c & a - b - c \\ 
a - b + c & a + 2 b & a - b \\
a - b - c & a - b & a + 2 b 
\end{array}
\right).
\end{equation}

In the present basis (symmetry basis) the charged lepton mass matrix is non-diagonal. 
Next, we make the transformation $M_l = U_L^{\dagger} m_l U_R$, which leads to a diagonal charged lepton mass matrix with 
\begin{equation}
 U_L=\frac{1}{\sqrt{3}}\left(
\begin{array}{ccc}
1 & 1 & 1 \\ 1 & \omega & \omega^2 \\ 1 & \omega^2 & \omega
\end{array}
\right)
\end{equation}

and $U_R = I$, where $I$ is a $3\times3$ unit matrix. In this basis, where the charged lepton mass matrix is diagonal, the neutrino mass matrix takes the following form:
\begin{equation}
 M_\nu \equiv U_L^{\dagger} m_{\nu} U_L^* = \left(
\begin{array}{ccc}
 3 a & -\frac{i c}{\sqrt{3}} &
   \frac{i c}{\sqrt{3}} \\
 -\frac{i c}{\sqrt{3}} & -\frac{2
   i c}{\sqrt{3}} & 3 b \\
 \frac{i c}{\sqrt{3}} & 3 b &
   \frac{2 i c}{\sqrt{3}}
\end{array}
\right).
\end{equation}
By redefining the elements of $M_{\nu}$ in above equation, we obtain the following form of the neutrino mass matrix in the flavor basis:
\begin{equation}
 M_\nu = \left(
\begin{array}{ccc}
x & y & -y \\ y & 2 y & z \\ -y & z & -2 y
\end{array}
\right).
\end{equation}
Since the above neutrino mass matrix is realized at the seesaw scale, there are also renormalization group corrections which need to be incorporated at the electroweak scale. However, these details are beyond the scope of present work.\\
\section{Conclusion}
In conclusion, we have presented a new form neutrino mass matrix with its characteristic predictions for neutrino masses, mixing angles, and CP violating phases. The atmospheric neutrino mixing angle $\theta_{23}$ is predicted to be near maximal. Although, the CP phase $\delta$ is allowed in its full range, it's highly correlated with the value of $\theta_{23}$. The effective mass $|M_{ee}|$ is also highly correlated with $\theta_{23}$. We obtain a lower bound  $|M_{ee}| > 0.046$ eV for the inverted neutrino mass ordering.  We also demonstrate how such a form of the neutrino mass matrix can be realized using the non-Abelian discrete symmetry $A_4$.

\renewcommand{\theequation}{A-\arabic{equation}}
\setcounter{equation}{0}
\section*{A \ \ The Group $A_4$}
$A_4$ has four inequivalent irreducible representations (IRs) which are three singlets \textbf{1}, $\textbf{1}^\prime$, and $\textbf{1}^{\prime\prime}$, and one triplet \textbf{3}. The group $A_4$ is generated by two generators $S$ and $T$ such that 
\begin{equation}
S^2 = T^3 = (S T)^3 = 1.
\end{equation}
The one dimensional unitary IRs are
\begin{equation}
\textbf{1} \  S = 1 \ \ T = 1, \ \
\textbf{1}^{\prime} \  S = 1 \ \ T = \omega, 
\textbf{1}^{\prime\prime} \  S = 1 \ \ T = \omega^2.
\end{equation}
The three dimensional unitary IR in the $S$ diagonal basis is 
 \begin{equation}
 S = \left(
\begin{array}{ccc}
1 & 0 & 0 \\ 0 & -1 & 0 \\ 0& 0 & -1
\end{array}
\right), \ \ T = \left(
\begin{array}{ccc}
0 & 1 & 0 \\ 0 & 0 & 1 \\ 1& 0 & 0
\end{array}
\right).
\end{equation}
The multiplication rules of the IRs are as follows
\begin{equation}
\textbf{1}^\prime \otimes \textbf{1}^\prime = \textbf{1}^{\prime \prime}, \ \textbf{1}^{\prime \prime} \otimes \textbf{1}^{\prime \prime} = \textbf{1}^{\prime}, \ \textbf{1}^{\prime} \otimes \textbf{1}^{\prime \prime} = \textbf{1}.
\end{equation}
The product of two $\textbf{3}$'s gives 
\begin{equation}
\textbf{3} \otimes \textbf{3} = \textbf{1} \oplus \textbf{1}^\prime \oplus \textbf{1}^{\prime \prime} \oplus \textbf{3}_s \oplus \textbf{3}_a,
\end{equation}
where $s$, $a$ denote the symmetric, anti-symmetric products, respectively. 
Let $(x_1, x_2, x_3)$ and $(y_1, y_2, y_3)$ denote 
the basis vectors of two $\textbf{3}$'s. IRs obtained from 
their products are
\begin{align}
(\textbf{3}\otimes\textbf{3})_{\textbf{1}} & = x_1 y_1 + x_2 y_2 + x_3 y_3    \\
(\textbf{3}\otimes\textbf{3})_{\textbf{1}^\prime} & = x_1 y_1 + \omega^2 x_2 y_2 + \omega x_3 y_3   \\
(\textbf{3}\otimes\textbf{3})_{\textbf{1}^{\prime \prime}} & = x_1 y_1 + \omega x_2 y_2 + \omega^2 x_3 y_3   \\
(\textbf{3}\otimes\textbf{3})_{\textbf{3}_s} & = (x_2 y_3 + x_3 y_2, x_3 y_1 + x_1 y_3, x_1 y_2 + x_2 y_1)   \\
(\textbf{3}\otimes\textbf{3})_{\textbf{3}_a} & = (x_2 y_3 - x_3 y_2, x_3 y_1 - x_1 y_3, x_1 y_2 - x_2 y_1).
\end{align}

\end{document}